*Review*

# Social Capital Contributions to Food Security: A Comprehensive Literature Review

**Saeed Nosratabadi[1], Nesrine Khazami[1], Marwa Ben Abdallah[1], Zoltan Lackner[2], Shahab S. Band[3,*], Amir Mosavi[4,5,6,7,8,*], and Csaba Mako[9]**

[1] Doctoral School of Management and Business Administration, Szent Istvan University, 2100 Godollo, Hungary; saeed.nosratabadi@phd.uni-szie.hu (S.N.); khazami.nesrine@phd.uni-szie.hu (N.K.); marwa.ben.abdallah@phd.uni-szie.hu (M.B.A.)

[2] Department of Food Economics, Faculty of Food Science, Szent Istvan University, Villanyi str. 29-43, 1118 Budapest, Hungary; lakner.zoltan@etk.szie.hu

[3] Future Technology Research Center, National Yunlin University of Science and Technology, Douliou, Yunlin 64002, Taiwan

[4] Faculty of Civil Engineering, Technische Universität Dresden, 01069 Dresden, Germany

[5] Thuringian Institute of Sustainability and Climate Protection, 07743 Jena, Germany

[6] School of Economics and Business, Norwegian University of Life Sciences, 1430 Ås, Norway

[7] John von Neumann Faculty of Informatics, Obuda University, 1034 Budapest, Hungary

[8] School of the Built Environment, Oxford Brookes University, Oxford OX3 0BP, UK

[9] Department of Public Management and Information Technology, National University of Public Services, 1083 Budapest, Hungary; mako.csaba@tk.mta.hu

* Correspondence: shamshirbands@yuntech.edu.tw (S.S.B.); amir.mosavi@mailbox.tu-dresden.de (A.M.)

**Abstract:** Social capital creates a synergy that benefits all members of a community. This review examines how social capital contributes to the food security of communities. A systematic literature review, based on Prisma, is designed to provide a state-of-the-art review on capacity social capital in this realm. The output of this method led to finding 39 related articles. Studying these articles illustrates that social capital improves food security through two mechanisms of knowledge sharing and product sharing (i.e., sharing food products). It reveals that social capital through improving the food security pillars (i.e., food availability, food accessibility, food utilization, and food system stability) affects food security. In other words, the interaction among the community members results in sharing food products and information among community members, which facilitates food availability and access to food. There are many shreds of evidence in the literature that sharing food and food products among the community member decreases household food security and provides healthy nutrition to vulnerable families and improves the food utilization pillar of food security. It is also disclosed that belonging to the social networks increases the community members' resilience and decreases the community's vulnerability that subsequently strengthens the stability of a food system. This study contributes to the common literature on food security and social capital by providing a conceptual model based on the literature. In addition to researchers, policymakers can use this study's findings to provide solutions to address food insecurity problems.

**Introduction**

Food security can be realized by accessing a balanced diet and essential nutrition for a healthy life [1]. Achieving food security has become one of the most important goals of governments and international organizations. The number of people exposed to food insecurity is on the rise at a fast-paced. Vulnerability to food security has significantly risen from 1693.3 million, in 2014, to 2013.8 million in 2018 [2]. It is estimated that around 704.3 million people had faced severe food insecurity in 2018 [2]. Rapid population growth, changing lifestyles, and international institutions' efforts to alleviate poverty are factors that have fueled the growing demand for food. It is estimated that the world's population will exceed 10 billion by 2050 [3], while the number of undernourished people has been increasing since 2015 [2] as it has reached 815 million in 2018 [4].

According to Schmidhuber and Tubiello [1], there is food security when all human beings have access to the nutrition and food preferences needed for a healthy life. To measure food security, Ruane and Sonnino [5] consider four criteria: Food availability, food accessibility, food utilization, and food system stability. As food availability indicates that high quality and nutritious food should be available in a region, regardless of whether it is produced or processed locally or internationally. Food access means that people need to be able to access food both physically and economically. Food utilization refers to the fact that all age groups should have access to healthy food that includes proper nutrition to live a healthy life. Ultimately, food system stability explains a system that provides enough food to the community and is also resilient to economic and climate shocks.

On the other hand, there is ample evidence that climate change has had a negative impact on crops and food productions. Climate change has led to droughts that have dramatically diminished agricultural yields as temperatures rise and changes in precipitation regimes, and it is expected that the impacts will be even exacerbated by 2050 [6]. Several solutions and factors had been recently proposed to address food security. For instance, land management [7], advanced biotechnologies [8], and water management [9] were used to improve agricultural efficiency. Increasing the financial statement of households (e.g., Reference [10]) through new policies to raise the level of education [11] has also shown promising results. Interdisciplinary research has provided valuable findings in the fight against food insecurity, one of which is social capital [12,13].

Social capital can contribute to food security through the synergy that is created from the interrelationship among community members at every stage of the food supply chain from production to consumption. In fact, social capital is the benefits that society derives from the interaction between different networks and groups [14]. Interpersonal relationships within social networks provide benefits to individuals through trust and social support (i.e., bonding capital). On the other hand, the interrelationship between these social networks will bring benefits to each of these networks by exchanging information, resources, and support (i.e., bridging capital) [15]. In the literature, the total benefits that individuals receive from membership in social groups and the benefits that society and each of these groups get from interacting with one other are called social capital [15]. Social capital, the synergy resulting from members of a community's interactions, brings benefits to community members and is a tool that members of a community can use as a solution to problems, such as food security. Kansanga et al. [14] believe that social capital is the resources that are created in human networks with common norms that facilitate social transactions and facilitate achieving the common goals of society for members. Social capital is identified through social organization characteristics, such as trust, norms, and networks. Social capital has a multidimensional character. According to one of the most comprehensive empirical study of the World Bank, it is composed of the following six dimensions [16]: (1) groups and networks, (2) trust and solidarity, (3) collective actions and cooperation, (4) information and communication, (5) social cohesion and inclusion, and (6) empowerment and political action. Due to its multi-dimensional character "as a topic, then, social capital tends lends itself to a mixed-methods research approach. Employing both qualitative and quantitative methods allows researchers to uncover the links between different dimensions of social capital construct a more comprehensive picture off the structure and perceptions of social capital" [16]. Trust is one of the most important dimensions of the social capital, to illustrate its core "social cement" function in the society and institutions, it is worth to quote the founder of the well-known Chinese global high-tech firm: "As Jack Ma, founder of Alibaba, famously said, "when you trust, everything is simple. If you don't trust, things get complicated" [17]. Social capital combines beliefs, rules of behavior, and interpersonal links,
without giving reasons why such a comprehensive definition is useful to our understanding of the social world. Several authors have defined social capital in an even more inclusive way, where even attitudes towards others, for example, appear: Social capital outlines trust, concern for others, desire to live according to the norms of one's society and to punish those who oppose it [18]. The literature proves that social capital and the synergy resulted from interactions among community members improve food security status, both directly and indirectly. For example, Martin et al. [12] Show that social capital reduces hunger, and in another study, Sseguya [19] shows that social capital has improved food security in southern Uganda. However, these studies are fragmental. Therefore, the present study aimed to provide a platform in which the results of as much as possible studies are collected. It is tried to present a comprehensive and integrated picture of how social capital affects food security. In other words, this study's main contribution is to show how social capital can improve food security.

There are many studies that have used the benefits of social capital to achieve the goals of food security. However, these studies are fragmental, and there is no complete picture of how social capital can contribute to food security in the literature. Some studies have also used social capital to address food security, but have not directly referred to it as a solution to social capital for food supply. To provide a clear understanding of the solutions that the research of social capital has so far provided for food security, the present study intends to bridge this gap in the literature by systematically reviewing the literature. Therefore, the main research question that the present study is addressing is how does social capital improves food security? In other words, by answering this question, this study tends to provide a theoretical framework illustrating how social capital contributes to the improvement of food security in different stages of the food supply chain. To this end, this study first identifies and reviews the published documents in these two areas (i.e., food security and social capital), and accordingly, corresponding hypotheses are presented for designing the theoretical framework. A food supply chain (FSC) is a network of actors who deliver food from farms to final consumers in stages. The actors of the FSC are farmers, processors, distributors, retailers, and consumers (e.g., References [20,21]). Where farmers harvest the primary crop, processors, and pack the final products, distributors deliver the final products to retailers, and retailers are the final destination where the final products are delivered to consumers [22]. The present study uses a similar approach to analyze the role of social capital in each of the food supply chain stages, which means that this study considers farmers, food processors, distributors, retailers, and consumers handling as the food supply chain stages. Classifying articles based on FSC steps first shows which stages of the supply chain the research has mostly focused on, and second, it shows what solutions are provided to improve food security at each stage of the FSC.

In Section 2, the methodology used in the present study and the path of selecting the reviewed articles is described in detail. In Section 3, the findings resulted from reviewing the literature are presented, and the contributions of social capital to food security were identified, and the corresponding hypotheses were presented to shape the conceptual model of the study. The conclusion is presented at the end, and remarks and recommendations are provided based on the study's contributions.

## 2. Materials and Methods

To ensure that the greatest number of articles are found in the common field of food security and social capital, the present study used the Prisma method. The Prisma method is a systematic literature review method constituting four steps: (1) Identification, (2) screening, (3) eligibility, (4) inclusion [23,24]. These four steps are designed in such a way that researchers can find the most relevant articles to the subject under study. A multidisciplinary perspective was adopted to ensure that the maximum number of related articles was reached, and searches of databases were not limited to any journals or any specific category of journals. This method's output is a database of relevant articles that are ready for quantitative and qualitative analysis. For the current study, at the identification step, 321 articles were found through initial searches in Thomson Reuters Web-of-Science (WoS) (148 out of 321 articles) and Elsevier Scopus databases (173 out of 321 articles). It is worth noting that the search among the mentioned databases took place until the end of September 2020. To this end, the keywords of social capital and food were searched simultaneously in the title, abstract, and keywords of these databases' articles. Four more articles were added in this step to the identified articles based on the cross-referencing process. During the screening step, the articles passed through two filters of checking duplicates and relevance. At this step, first, the duplicate articles which were found in both databases were deleted 185 individual articles went to the next screening step. In this step, the articles are evaluated based on the relevance of their titles and abstracts where, as a result, 101 articles are eliminated, and 84 articles went to the eligibility step. At the eligibility step, the full text of 84 articles was carefully read, and finally, 39 articles were included in the data of the present study

(the inclusion step). All the results presented in this study are based on the analysis of these 39 articles. Figure 1 shows a visual view of the process of implementing the Prisma systematic literature review method in this study.

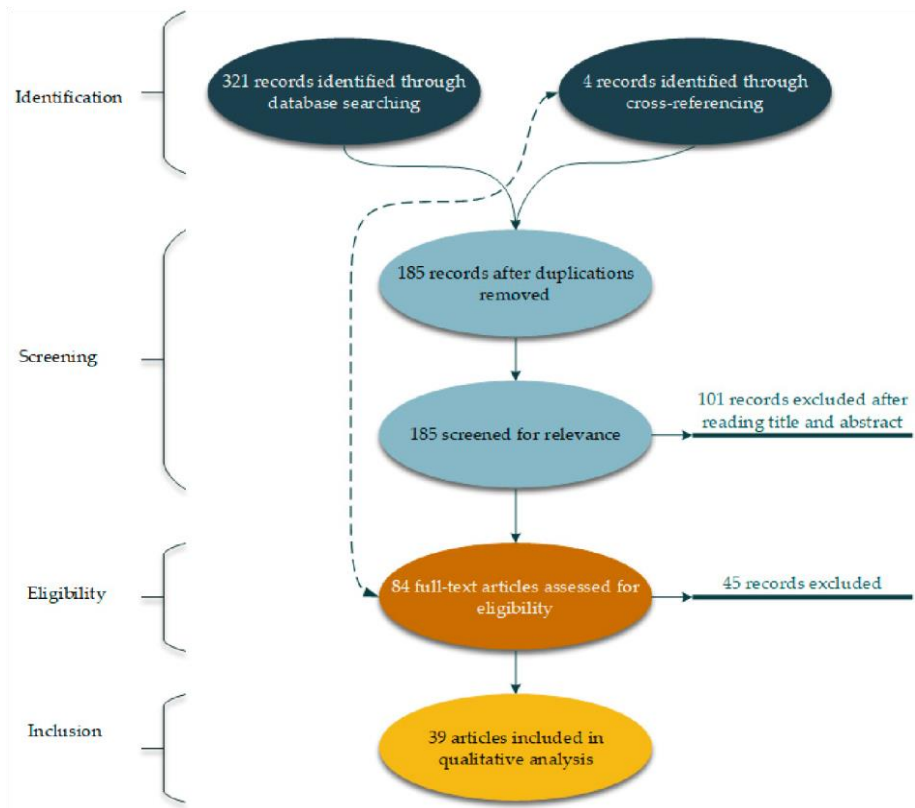

**Figure 1.** Diagram of the systematic selection of the study database.

For decades, food security has been the main concern of a large number of studies. Different studies have been tackled various sides of issues that affect food security. Social capital is one factor that could affect an area's food security level; thus, an essential number of papers have been issued exploring these relationships. Figure 2 points out the movement of the number of published papers between 1998 and 2020. It is highlighted that an increasing fluctuation shape was followed. A peak is registered in 2017, which explains the rise of the research interest in examining the link between social capital and food security.

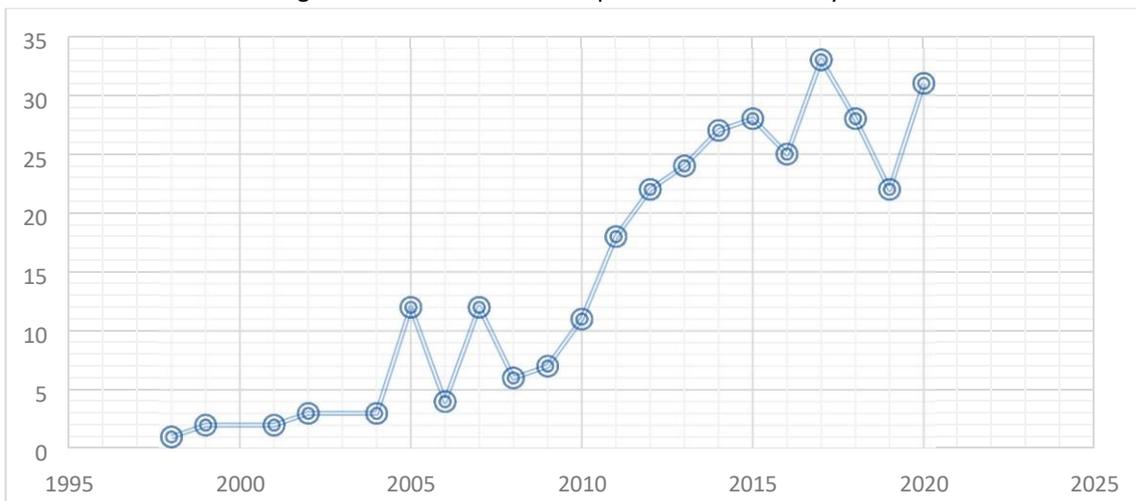

**Figure 2.** Trends of publications in the common area of food security and social capital from 1998 to 2020.

The published reviewed research documents employed three different research types. As indicated in Table 1, 21 research documents out of 39, more than half of the reviewed papers, have used a quantitative empirical research method. Qualitative empirical research with 15 articles is in second place, 38% of the total reviewed documents. The third type is the conceptual research method with three studies and constitutes only 7% of the reviewed papers.

**Table 1.** Documents based on research type.

| Explanation | Conceptual | Qualitative Empirical | Quantitative Empirical |
| --- | --- | --- | --- |

| Sources | Kiara [25]; Misselhorn [26]; Vitiello and Wolf-Powers [27]. | Chriest and Niles [28]; Furness and Gallaher [29]; Naughton, Deubel, and Mihelcic [30]; Pascoe and Howes [31]; Browne, Goncalo, Ximenes, Lopes, and Erskine [32]; Saint Ville, Hickey, and Phillip [33]; Olivier and Heinecken [34]; Helicke [35]; Kerr [36]; Mtika [37]; Kismul [38]; Gray et al. [39]; Porter and McIlvaine-Newsad [40]; Whitley [41]; Kaschula [42]. | Mwakiwa, Maparara, Tatsvarei, and Muzamhindo [43]; Lee et al. [44]; Sseguya, Mazur, and Flora [45]; Smith and Frankenberger [46]; Meador and Fritz [47]; Boubacar, Pelling, Barcena, and Montandon [48]; Wossen et al. [49]; Saint Ville et al. [50]; de Jalón, Iglesias, and Neumann [51]; Quetulio-Navarra, Frunt, and Niehof [52]; Rayamajhee and Bohara [53]; Kirkpatrick and Tarasuk [54]; Misselhorn [55]; Martin et al. [12]; Garrett and Leeds [56]; Chen, Wang, and Huang [57]; Gallaher et al. [13]; Nyikahadzoi et al. [58]; Bunch et al. [59]; Kaiser, Barnhart, and Huber-Krum [60]; Olarinde et al. [61]. |
|---|---|---|---|
| Number | 3 | 15 | 21 |

The articles reviewed in this study used different tools to collect data for reaching their research objectives. Figure 3 illustrates the data collection methods employed in the 39 research papers. The questionnaire was the most common data collection tool among the reviewed articles as it is employed with 15 publications. In addition, 13 papers used the interview as the data source of their research. Secondary database and Literature synthesis methods are performed within three publications. Case study and Ethnography with, respectively, four and one publications are other data collection methods applied among the articles. Figure 3 also shows that 90% of the total documents reviewed are original research papers, while review articles and book chapters with 6% and 4%, respectively, are other types of documents that were reviewed.

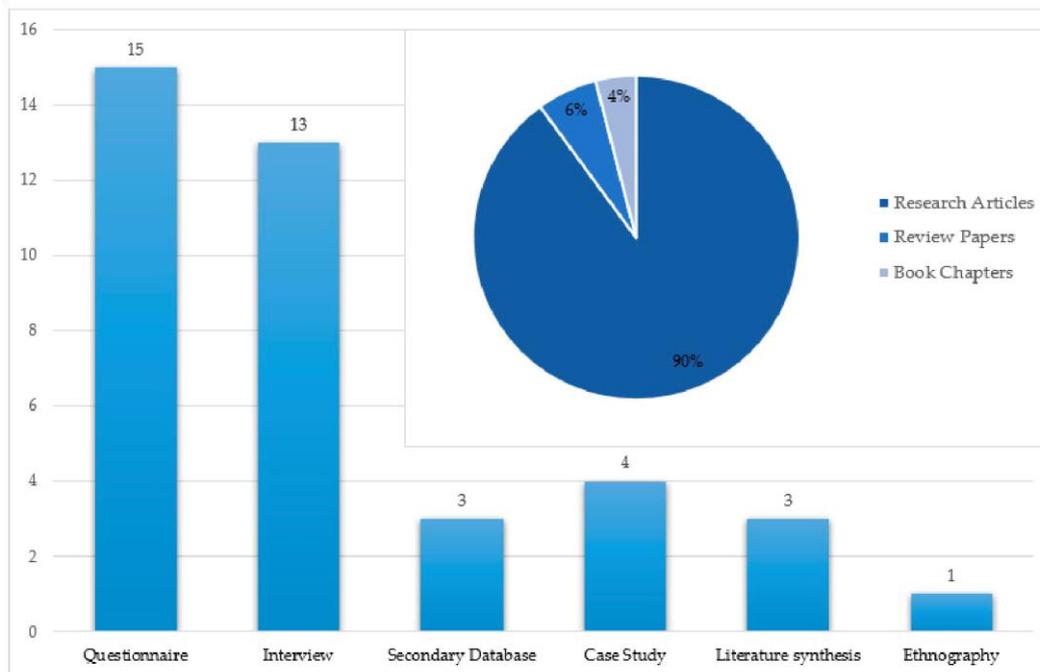

**Figure 3.** Data source and document types of reviewed articles in the current study.

Various journals cover food security and social capital issues simultaneously, and the journals with the largest contributions are listed in Table 2. The Food Security journal is on the head list with the highest paper number (6 papers), followed by the Global Environmental Change journal with three papers. Agriculture and Human Values, Development Southern Africa, Public Health Nutrition, and Rural Sociology journals are the journals with two published papers in this realm.

**Table 2.** Journals with the largest number of documents.

| Journal Name | Number of Publications |
|---|---|
| Food Security | 6 |
| Global Environmental Change | 3 |
| Agriculture and Human Values | 2 |
| Development Southern Africa | 2 |
| Public Health Nutrition | 2 |

Numerous countries have explored social capital and food security concepts. Within the 39 reviewed papers, 33 countries are identified. There are countries where food security and social capital have been studied more than once, which shows the importance of these two concepts in the respective countries. As highlighted in Figure 4, these articles are more concentrated in some countries than others. The blue color designs the USA, which has the highest number of publications, 14 papers. Indonesia and Kenya, presented in orange color, with three research papers, are other countries with the highest number of publications on these topics. Moreover, countries, such as Zimbabwe, Peru, Uganda, South Africa, and Malawi, have had two published papers in the common area of these two concepts, and they are presented in Figure 4 with the dark grey color. The countries with a single publication are scattered all over the world and presented in yellow color in Figure 4.

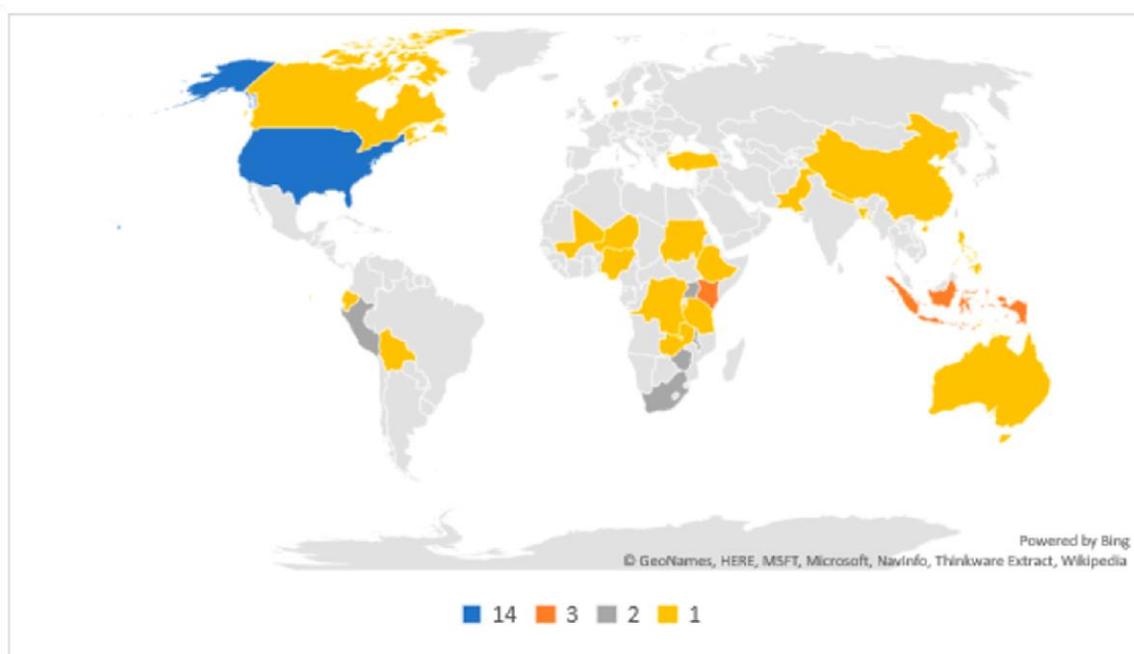

**Figure 4.** Distribution of publications on food security and social capital in different countries.

## 3. Results

The reviewed articles are classified based on their focus on the four pillars of food security (i.e., availability, accessibility, utilization, and food system stability) and the food supply chain actors (i.e., farmers, processors, distributors, retailers, and consumers). The finding revealed that social capital solutions for food security are mainly for two of the pillars of food security, i.e., food availability and food accessibility, as 77% of reviewed papers have focused on these two pillars. Sixteen one out of thirty-nine reviewed papers treats the availability side of food security. Examining the articles' focus on the food supply chain, it was found that 64% of the articles targeted farmers (see Table 3), and none of the articles have solutions for the food distributions. Table 3 also illustrates that most of the solutions presented in the literature are for food accessibility after food availability, where 14 of the 39 articles focused on this food security pillar. Six and three research papers then suggest solutions for food utilization, food system stability, respectively.

**Table 3.** Categorizing the articles based on their focus on food security strategies and on the food supply chain.

| Explanation | Food Availability | Food Accessibility | Food Utilization | Food System Stability |
|---|---|---|---|---|
| Farmers | Saint Ville et al. [50]; Kerr [13]; Saint Ville, Hickey, and Phillip [33]; Olivier and Heinecken [34]; de Jalón, Iglesias, Neumann [51]; Browne et al. [32]; Helicke [35]; Porter and McIlvaine-Newsad [40]; Garrett and Leeds [56]; Gray et al. [39]; Mwakiwa et al. [43]; Nyikahadzoi et al. [58]; Pascoe and Howes [31]; Vitiello and Wolf-Powers [27]; Gallaher et al. [13]. | Olarinde et al. [61]; Kiara [25]; Furness and Gallaher [29]; Lee et al. [44]; Sseguya, Mazur, and Flora [45]; Meador and Fritz [47]; Boubacar et al. [48]; Quetulio-Navarra et al. [52]. | | Chen et al. [57]; Wossen et al. [49]. |

| | | | |
|---|---|---|---|
| Processors | Naughton, Deubel, Mihelcic [30]. | | |
| Retailers | | Smith and Frankenberger [46]. | Kirkpatrick and Tarasuk [54]. |
| Consumers | | Misselhorn [26]; Whitley [41]; Kaschula [42]; Bunch et al. [59]; Kaiser et al. [60]. | Misselhorn [55]; Mtika [37]; Kismul [38]; Rayamajhee and Bohara [53]; Martin et al. [12]. | Chriest and Niles [28]. |

Figure 5A discloses that the vast majority of the articles focus on the farmers, where the food products are produced. Moreover, 11 out of 39 articles have concentrated on food security problems in the consumption section. Figure 5B also shows that food availability and food accessibility with 30 out of 39 articles are the food security pillars that have the most attention in the common literature of food security and social capital.

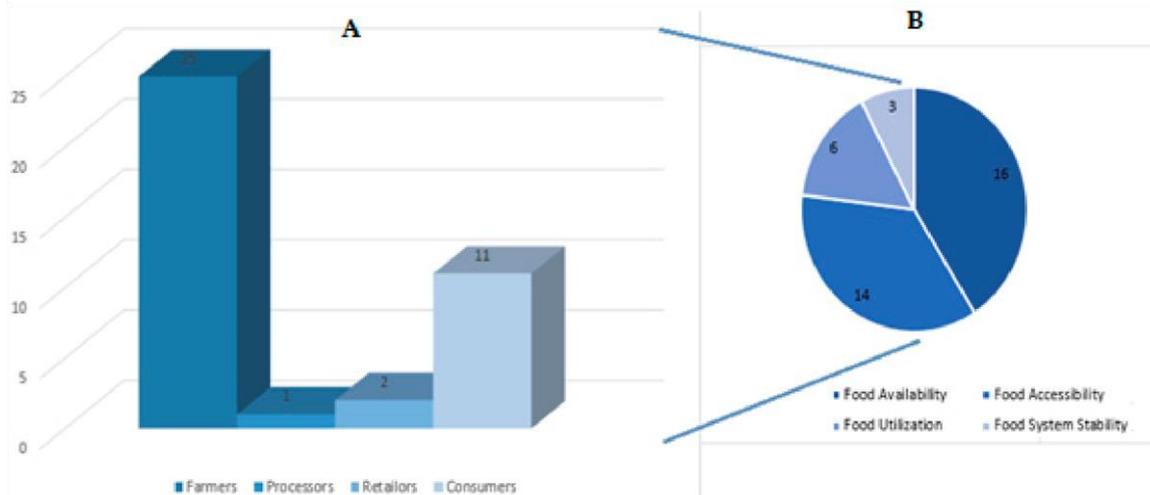

**Figure 5.** Classifying the reviewed articles based on their focus on (**A**) the food supply chain and (**B**) the food security pillars.

Network interconnections, interactions, and social norms within a community are called social capital [14]. Therefore, this study's main idea was to investigate how the network interconnections, interactions, and social norms of a community can improve food security. There have been many studies in this area, and these studies have sometimes yielded conflicting results. Hopkins and Holban [62], for example, within their investigation applied to the Applachian community, found that food insecurity is not related to social capital and food security in this region is strongly associated
with nutrition utilization and economic power of the community members (i.e., food accessibility) [62]. In another study in Saptosari Gunungkidul, Indonesia, it was also revealed that although there is a positive relationship between social capital and food security, this relationship is not significant [63]. Moreover, Olivier and Heinecken [34] found that urban agriculture, as an activity that enhances food security, promotes social capital in the community. In contrast, Dean et al. [64] show that social capital has a positive and significant relationship with food security. In a study of people between the ages of 50 and 59 in a rural area of central Texas, Dean et al. [64] found that people had less food security
where social capital and social interactions were low. The following are solutions for improving food security through social capital based on their contribution to each of the pillars of food security.

*3.1. Food Availability*

Ample evidence has been found in the literature that social capital can improve food security by improving food availability in various ways. Food availability, indeed, refers to the availability of adequate high-quality food in a region. The findings of a case study in Zimbabwe show that social capital, along with innovative technologies and market information, is a prerequisite for food availability, and their contribution to improving food security is higher than farm research initiatives [58]. A study conducted by Kerr [36] disclosed that Malawian smallholder farmers are running out of food in some months of the year and have to do Ganye. In fact, Kerr [36] argues that Ganye is a kind of piecework labor working on the others' farms in which the laborers are compensated by food or cash. Although Kerr [36] believes that Ganye alone is a very weak form of social capital, this social activity increases food availability [36]. Browne et al. [32] Show that rituals have a significant effect on food security and food availability in Timor-Leste. They argue that ritual support the production of some agricultural products (such as rice and maize), and these products are widely used in most ritual ceremonies. In addition, Browne et al. [32] believe that the act of sacrificing animals in rituals is another influential factor in increasing the production and the population of these animals. De Jalon et al. [51] argue that social capital can contribute to food security by increasing farmers' capacity to adapt to climate change. De Jalon et al. [51] show that the exchange of knowledge gained through interactions among the smallholder farms' community in sub-Saharan Africa has a positive and significant impact on farm-management practices, using fertilizers and agrochemicals, management of pest and crop, soil conservation practices, applying appropriate irrigation systems, utilizing more resistant crop varieties, each of which is a strategy for adapting to climate change.

Saint Ville et al. [50] investigate social capital's role in supporting innovation in smallholder farming systems. They figured out that there are strong social connections among the smallholder farmers in the Caribbean where such social connections exchange and make available the information among the farmers. In such a social network, various support resources that promote food availability are maintained [50]. Saint Ville et al. [33] argue that stakeholder interactions affect national food security policy. The result of their study reveals that there is a need to provide supportive conditions, such as alignment of different forms of social capital built based on the interactions among stakeholders, and trust and knowledge exchange among stakeholders in the policy network, for aggregating multi-stakeholder interactions in the development of Saint Lucia's National Agricultural Policy. To conserve the availability of two traditional varieties of wheat in Turkey, Helicke [35] found that social capital is an essential factor to connect small producers, national and external actors, like lack of social networks, threats the seed exchange connections [35]. In other words, they imply that social capital contributes to food availability through facilitating seed exchange. Membership of food producers in related communities can have benefits for producers and improve the quality of food security. In this regard, Naughton et al. 2017 [30] argue that the women producing shea butter and belonging to the shea butter organization were more motivated and more committed to producing shea butter, and they could access more markets and have a higher ability to negotiate revenues.

Makita et al. [43] believe that social capital plays a sustainable role in producing foods for households. They articulate that working on community gardens boosts social capital and will be resulted in the food supply while managing resources. Social networks, such as gardening activity, captivate people to maintain their food security. Not only is food security the goal of people participating in common gardening, but also it creates an opportunity for divertissement where people could gather, share information and enhance the social networks in rural areas [40]. Assessing the different determinants of community gardening in Philadelphia, it is highlighted that community gardening boosts food security and social capital [56]. In Saint-Jose, California, it is revealed that home gardens get a beneficial link with social capital [39]. Studying home gardening in Saint-Jose, California, Grey et al. [39] found that home gardening is a linkage between food security and social capital. Because they argue that in addition to providing food, food gardening increases bonding social capital through the sharing of knowledge and products. There are other studies in the literature with similar results, but these studies have used different terminology, Pascoe and Michael Howes [31], for example, use community gardening, Gallaher et al. [13] use sack gardening and urban agriculture, and Vitiello and Wolf-Powers [27] use urban agriculture terms. All these studies imply that social capital improves food availability primarily through knowledge and product sharing, which in turn improves food security. Hence, two hypotheses that can be extracted from these articles are:

**Hypothesis 1 (H1).** *Social capital improves food availability through sharing knowledge among members of a community.*

**Hypothesis 2 (H2).** *Social capital improves food availability through sharing products among members of a community.*

*3.2. Food Accessibility*

According to the definition, food accessibility refers to people's ability to access physically and economically to food. Furness and Gallaher [29], Lee et al. [44], Boubacar et al. [48], Smith and Frankenberger [46], Kaschula [42], and Quetulio-Navarra et al. [52], for instance, show that interaction among members of a community will result in the exchange of food products among them and increases the food accessibility in the region.

A case study in Rockford, Illinois, Furness, and Gallaher [29] evaluates the relationship between community gardening and food accessibility. They explain the magnitude of benefits gained by both the community and non-community members of these gardens. Because most of the products in these gardens go into food pantries and are accessible to others [29]. The results of a study by Olarinde et al. [61] reveal that Nigerian Cassava farming households who belonged to more social networks had higher opportunities to access food. Lee et al. [44] investigated the relationship between household food security and social connectedness among peri-urban Peruvian Amazonian communities. They found out that those households have more social connectedness within and outside the community reported less food insecurity, especially as the food was regularly shared among the members. Bunch et al. [59] illustrate that social capital increases the households' food security in Greater Bahr el Ghazal and the Equatorias in South Sudan. Boubacar et al. [48] developed a tool to assess household resilience in Niger. Their research shows that social capital can improve families' resilience exposed to floods and subsequently increase access to food in the area. In a study of households in northern Bangladesh, Smith, and Frankenberger [46] found out that social capital could increase food security by increasing households' resilience to natural shocks, such as floods. They show that increasing resilience in flood-prone areas increases the number of months that households can access enough food. Kaiser et al. [60] found that using social networks provides food for US households. In other words, they claim that belonging to social communities facilitates access to food for households. In a study of households in KwaZulu-Natal, South Africa, Kaschula [42] found that families with adults at risk for chronic illness were more likely to have food donated by community networks. However, these families have also been exposed to food insecurity. However, Kaschula [42] found that families with Acquired Immunodeficiency Syndrome (AIDS)/Human Immunodeficiency Viruses (HIV) have not benefited from these donated foods because of social labels. Quetulio-Navarra et al. [52] showed that culturally underpinned food exchange practices in Central Java, Indonesia, facilitated the food accessibility among the households that have subsequently led to children's food security five years old in this area. All these studies were implying the importance of sharing food products among community members in improving access to food. Therefore, the following hypothesis can be deducted:

**Hypothesis 3 (H3).** *Social capital improves food accessibility through sharing products among members of a community.*

There is evidence in the literature that social capital can contribute to food accessibility by sharing information among members of a community. For example, Whitley [41] defines "food deserts," a place in which there is no grocery store or only one with limited and expensive food items. The results of his study show that people with less social connection reported more problems in having access to food, and in contrast, those with more social ties reported no problems accessing food. The results of the study of Sseguya et al. [45] also disclose that social capital is linked positively with food access. Studying households' insecurity in rural Uganda, Sseguya et al. [45] illustrate that social capital, through increasing access to information and sharing and shaping norms and mutual trust, improves food access among the households. By analyzing the studies that took place in African, Misselhorn [26] found that low social capital among the communities increased poverty and conflict among the community and reduced human health, which in turn led to less accessibility to healthy foods among members of the community. Misselhorn [26] believes that social capital can improve food security and by building the social resilience of the community. Studying the participatory community planning approach to agricultural extension in Kenya, Kiara [25] shows that increasing smallholder farmers' interactions through public extension processes have increased women's, youths', poor, and vulnerable groups' engagement in generating information and providing solutions to increase access to food. In addition, such community interactions promoted smallholder farmers from subsistence to business farming that contributes to the food security of the region as well. Microfinance organizations in Uganda are trying to improve food security and reduce poverty by providing farmers with financial resources. In a study of rural women in Uganda, Meador and Fritz [47] found that financial institutions can improve food accessibility by creating social capital by increasing the women's interactions in the community.

**Hypothesis 4 (H4).** *Social capital improves food accessibility by sharing knowledge among members of a community.*

*3.3. Food Utilization*

Among the articles reviewed, there are studies that show that the interactions of members of a community are inversely related to malnutrition. A study among 330 low-income households from Hartford, Connecticut, the USA, revealed that even when households' financial or food resources are limited, households with higher social status have lower food insecurity and are less likely to experience hunger [12]. A study by Misselhorn [55] in KwaZulu-Natal illustrated that social capital was directly related to food security, and factors (such as divorce, religious conflicts, cultural differences, and leadership conflicts) that undermined social capital ultimately led to food insecurity among members. By studying neighborhood characteristics and their relationship to household food security among low-income Toronto families, Kirkpatrick and Tarasuk [54] found that in neighborhoods with less perceived social capital, households faced the problem of lack of nutrients needed for a healthy life, in other words, these households were exposed to food insecurity. Rayamajhee and Bohara [53] consider involvement in voluntary associations as a measure of social capital and investigate the effect of such volunteering activities on food security among households in western Nepal. They found out participating in financial associations (such as micro-finance, insurance, trade, and business associations), as a volunteer has a direct impact on hunger mitigation, and participation in informational associations (such agriculture, water, forest groups) helps Nepali women to improve the nutritional quality of diets and the food utilization. According to Mtika [37], the AIDS epidemic among the Malawi rural households weakens the social immunity and social capital among the member. He found that weakening relationships among community members makes them very vulnerable and ultimately leads to food insecurity among rural households, who have serious difficulty providing healthy food and adequate nutrition for a healthy life. Studying the social context of severe child malnutrition in a rural area of the Democratic Republic of Congo, Kismul et al. [38] found solutions, including gbisa, to tackle food insecurity and child malnutrition. Gbisa, indeed, is inter-household cooperation in providing the labor force for the farms. In this form of cooperation, close kin and neighbors go to the others' farms to help the farmers, for example, inland clearing and timely weeding. In return, they receive enough food plus a kind of capital and even cash as compensation. These studies illustrate that social capital can improve food utilization through sharing products and information among members of a community. Therefore, the hypothesis can be deducted is as follow:

**Hypothesis 5 (H5).** *Social capital improves food utilization through sharing products among members of a community.*

*3.4. Food System Stability*

Food stability outlines that all people should have access to enough food all the time, regardless of any unforeseen risk (e.g., weather change, pandemic, economic crisis, etc.), which could prevent households from accessing food. Food insecurity is a problem of food abundance and infrastructure and includes the group member's food access and their alliance. Thus, each member within a group of people should access their common resources inside their territory. Chriest and Nilest [28] found that high social capital in the rural communities in the US can increase the resilience and adaptation of the community to extreme weather events, and it also increases food security to deal with problems caused by climate change shocks. Studying Ethiopian rural households, Wossen et al. [49] found that shocks, such as drought, market shocks (i.e., food price), and health shocks, in both household and rural levels, increased food insecurity. Their findings also reveal that membership in informal communities and organizations reduced these shocks' effects on household food security. Chen et al. [57] believe that government support policies to combat drought have made a significant contribution to stabilizing the food system and have been able to help improve food security in China by increasing agricultural production [57]. Social interactions that increase the resilience and eliminate the households' vulnerability will result in a more stable food system and a decline in food insecurity. Thus, the following hypothesis can be deducted: H6: *Social capital activities strengthen food system stability*. Reviewing 39 articles resulted in the formation of 6 hypotheses, which are shown in Figure 6. Figure 6 illustrates that social capital improves food availability and food accessibility through two mechanisms: Knowledge sharing and product sharing (i.e., food sharing). Food utilization, on the other hand, is developed only through a product-sharing mechanism. Hypothesis 6 in Figure 6 states that social capital increases the stability of the food system. In general, the social capital and interactions of the members of a society stabilize a food system by eliminating the members' vulnerability and increasing their resilience.

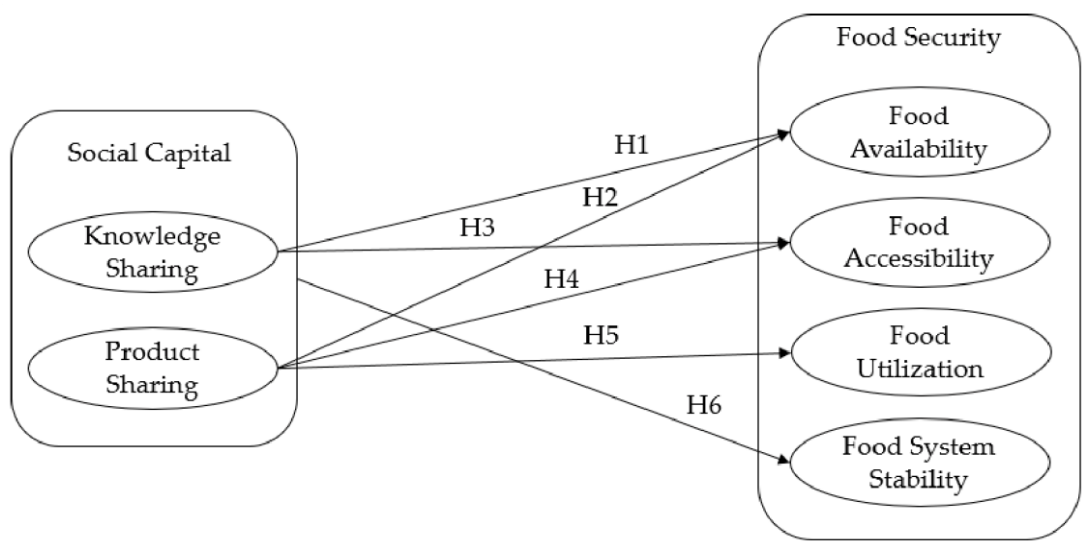

**Figure 6.** Conceptual model of the study; the effect of social capital on food security. H1: Social capital improves food availability through sharing knowledge among members of a community;.H2: Social capital improves food availability through sharing products among members of a community; H3: Social capital improves food accessibility through sharing products among members of a community; H4: Social capital improves food accessibility by sharing knowledge among members of a community; H5: Social capital improves food utilization through sharing products among members of a community; H6: Social capital activities strengthen food system stability.

The present study's findings are consistent with the existing literature, where Mastronardi, Giagnacovo, and Romagnoli [65] believe that community-based cooperatives increase community resilience to external shocks. On the other hand, Mastronardi and Romagnoli [66] believe that such cooperation and the formation of such communities will lead to society's development and sustainability. Based on the findings, members of a community can benefit from each other's support, leading to food security in various ways. It was found that these supports come in two general forms: (1) Through product sharing and (2) through information sharing. Product sharing is done in different ways. In ganye, for example, individuals in a community go to each other's farms as labor, and not only do they contribute to the production of crops, but they also receive food as compensation. As another example, social communities that distribute foods among vulnerable people exposed to food insecurity increase their food security status. Other studies have highlighted the importance of sharing knowledge and sharing experience in food security among community members. For example, Helicke [35] shows that creating a network between small farmers in Turkey not only led to the exchange of seeds, but also helped members of the group to share their experience and knowledge with other members of the group to find a market, which increases food availability in turn. Community gardening is another example in which community members share their experience and knowledge to produce food. There is ample evidence in the literature that community gardening improves food security for both members and non-members of a community.

## 4. Conclusions

Social capital, and the synergy from community members' interactions, improve food security status—both directly and indirectly. Therefore, the present study aimed to provide a platform in which the results of as much as possible studies are collected. It is tried to present a comprehensive and integrated picture of how social capital affects food security. In other words, this study's main contribution is to show how social capital can improve food security. A detailed review of 39 articles published in the field of food security and social capital found that social capital enhances food security through two mechanisms of knowledge sharing and product sharing (i.e., sharing food and food products). Findings show that sharing food and food products within a community facilitates food accessibility and reduces hunger. On the other hand, sharing knowledge and information among community members expands the farms' markets and increases community resilience to unexpected shocks. Moreover, it is also disclosed that community gardening plays an essential role in food security. Community gardening is a platform through which a social network is formed, and its members exchange experience and knowledge and cooperate in the production of and share of food among the community members, which ultimately enhances food availability and food accessibility. Therefore, it is suggested that actors at different stages of FSC form networks and communities, such as community-based cooperatives, because they can contribute to both food security by increasing food availability and increasing community resilience to extreme events. Hence, the policymakers, NGOs, the other initiatives that can form social networks can benefit from the finding of the study to develop policy and necessary actions to build social capitals for dealing with food security issues. These results also present a basis for future research. Future research can apply a quantitative research methodology to test the hypotheses presented in this study and test this study's conceptual model.

**Author Contributions:** Conceptualization, S.N. and A.M., N.K.; methodology, S.N., S.S.B.; validation, A.M., Z.L.; investigation, S.N.; data analysis, S.N., N.K., M.B.A.; writing—original draft preparation, S.N., N.K., and M.B.A.; writing—review and editing, S.N., N.K., M.B.A., Z.L., C.M., S.S.B., and A.M.; visualization, A.M., and S.N.; supervision, C.M., and Z.L. All authors have read and agreed to the published version of the manuscript. **Funding:** This research has been supported by the Alexander Von Humboldt foundation **Acknowledgments:** We acknowledge the support of the Alexander Von Humboldt Foundation.

**Conflicts of Interest:** The authors declare no conflict of interest.